\documentclass{ar-1col}
\usepackage[numbers]{natbib}
\usepackage{graphicx}
\usepackage{amsmath}
\usepackage{subfigure}
\newcommand{\beq}{\begin{equation}}
\newcommand{\eeq}{\end{equation}}
\newcommand{\bk}{{{\bf{k}}}}

\newcommand{\bA}{{\bf{A}}}
\newcommand{\bB}{{\bf{B}}}
\newcommand{\bE}{{\bf{E}}}

\newcommand{\bq}{{\bf{q}}}

\newcommand{\bb}{{\bf{b}}}

\newcommand{\bj}{{\bf j}}

\newcommand{\beqa}{\begin{eqnarray}}
\newcommand{\eeqa}{\end{eqnarray}}
\newcommand{\pdg}{{\vphantom \dag}}
\newcommand{\dg}{{\dag}}
\newcommand{\bnabla}{{\boldsymbol \nabla}} 
\newcommand{\bsigma}{{\boldsymbol \sigma}}
\newcommand{\bgamma}{{\boldsymbol \gamma}}

\newcommand{\bOmega}{{\boldsymbol \Omega}}
\newcommand{\bGamma}{{\boldsymbol \Gamma}}

\jname{Xxxx. Xxx. Xxx. Xxx.}
\jvol{AA}
\jyear{YYYY}
\doi{10.1146/((please add article doi))}
\begin{document}

\markboth{A.A. Burkov}{Weyl Metals}

\title{Weyl Metals}

\author{A.A. Burkov
\affil{Department of Physics and Astronomy, University of Waterloo, Waterloo, Ontario N2L 3G1, Canada; email: aburkov@uwaterloo.ca}}

\begin{abstract}
Weyl metal is the first example of a conducting material with a nontrivial electronic structure topology, making 
it distinct from an ordinary metal. 
Unlike in insulators, the nontrivial topology is not related to invariants, associated with completely filled bands, but 
with ones, associated with the Fermi surface. 
The Fermi surface of a topological metal consists of disconnected sheets, each enclosing a Weyl node, 
which is a point of contact between two nondegenerate bands. 
Such a point contact acts as a source of Berry curvature, or a magnetic monopole in momentum space. 
Its charge, or the flux of the Berry curvature through the enclosing Fermi surface sheet, is a topological invariant. 
We review the current state of this rapidly growing field, with a focus on bulk transport phenomena in topological metals. 
\end{abstract}

\maketitle

\tableofcontents

\section{INTRODUCTION}
\label{sec:1}
A significant part of the modern condensed matter physics deals with macroscopic quantum phenomena, that is 
direct manifestations of quantum mechanics, which naturally governs the behavior of matter on the atomic and subatomic scales, 
on macroscopic scales, which are, roughly speaking, the size scales discernible by the naked human eye. 
In many instances, e.g. in the case of perhaps the most well-known macroscopic quantum effect, the superconductivity, the underlying source of such phenomena is electron-electron interactions, which 
make the electrons behave cooperatively, leading to macroscopic quantum coherence. 
Interactions are not the only source of the macroscopic quantum coherent behavior, however. 
Integer quantum Hall effect (IQHE), which manifests in a precisely quantized transverse resistivity of a two-dimensional 
electron gas (2DEG) system in a strong perpendicular magnetic field, results from the interplay of two purely quantum phenomena, 
which do not involve interactions: localization of electrons in a random impurity potential and nontrivial topology of the single-electron eigenstates in the presence of magnetic field. The nontrivial topology leads to delocalized metallic edge states in an otherwise insulating sample, which is what ultimately results in the quantized Hall effect. 

It is now well-understood that macroscopic quantum behavior, brought about by the nontrivial electronic structure topology, is not limited, as initially thought, to 2DEG in a strong perpendicular magnetic field. 
This new understanding started to emerge from the pioneering work of Haldane~\cite{Haldane88} and developed fully 
after the discovery of Topological Insulators (TI)~\cite{Hasan10,Qi11} and the tremendous amount of work that followed.  
The most recent new development in this field is the realization that not only insulators, but also metals may be topological. 
While partially anticipated in earlier work~\cite{Volovik03,Haldane04,Volovik07,Murakami07}, this idea was firmly established after 
the theoretical and later experimental discovery of Weyl and Dirac semimetals~\cite{Wan11,Ran11,Burkov11-1,Burkov11-2,Xu11,Kane12,Fang12,Fang13,HasanTaAs,Neupane14,DingTaAs2,DingTaAs,Lu15}. 

In a metal, nearly everything of observable consequence happens on the Fermi surface. 
Thus, in order for a metal to be ``topological", the corresponding momentum space invariant needs to be defined on the Fermi surface. 
This is in contrast to TI, where topological invariants have to do with completely filled bands and thus involve all states 
in the first Brillouin zone (BZ). 
As it happens, the only topological invariant, that may be defined on the Fermi surface, and that leads to observable 
consequences, is the flux of what is known as the Berry curvature through the two-dimensional (2D) Fermi surface of a 
three-dimensional (3D) metal. 

Berry curvature is a close analog of the magnetic field, but is defined in momentum rather than real space.
The Hamiltonian of noninteracting electrons in a crystal is a matrix $H(\bk)$, whose elements are labelled by the atomic 
orbital and spin quantum numbers and that depends in the crystal momentum $\bk$ as a parameter.
The electronic structure is defined by the eigenvalues and eigenvectors of the equation
\beq
\label{eq:1}
H(\bk) |u(\bk) \rangle = \epsilon(\bk) |u(\bk)\rangle. 
\eeq
Consider an overlap between a particular eigenvector taken at point $\bk$ and the same eigenvector 
taken at nearby point in momentum space $\bk + \delta \bk$
\beq
\label{eq:2}
\langle u(\bk) | u(\bk +\delta \bk) \rangle \approx 1 + \delta \bk \cdot \langle u(\bk) | \bnabla_\bk | u(\bk) \rangle 
\approx e^{i \bA(\bk) \cdot \delta \bk}, 
\eeq 
where $\bA(\bk) = - i \langle u(\bk) | \bnabla_\bk | u(\bk) \rangle$ is called the Berry connection. 
The curl of the Berry connection 
\beq
\label{eq:3}
\bOmega(\bk) = \boldsymbol \nabla_\bk \times \bA(\bk), 
\eeq
 is the Berry curvature. 
 When the Berry curvature is not zero, the phase $\bA(\bk) \cdot \delta \bk$ can not be eliminated  by a gauge transformation 
 and has observable consequences. 

Suppose we have a Fermi surface sheet, with a nonzero flux of the Berry curvature through it. 
Gauss' theorem implies that there must then be a point source or sink of the Berry curvature, enclosed by this Fermi surface. 
In the vicinity of this point, the Berry curvature takes, up to trivial rescaling of the crystal momentum components, a universal form 
\beq
\label{eq:4}
\bOmega(\bk) = \pm \frac{\bk}{2 k^3}, 
\eeq
such that 
\beq
\label{eq:5}
C = \frac{1}{2 \pi} \int \bOmega(\bk) \cdot d {\bf S} = \pm 1. 
\eeq
$C$ is the Chern number, that is associated with any 2D Fermi surface sheet, enclosing a point source or sink of the Berry curvature and may be regarded as the ``topological charge" of this point. 
The sign depends on whether the point is a source or a sink. 

A point source or a sink of the Berry curvature is clearly some sort of a singularity of the electronic structure. 
As we will see explicitly later, this singularity turns out to be a point of degeneracy between two bands. 
Such degeneracy points are called Weyl nodes~\cite{Wan11}. 
The name arises from the following important observation. 
Just as the Berry curvature in vicinity of the degeneracy point takes a universal form of the magnetic field of a point
monopole~\eqref{eq:4}, the band Hamiltonian itself, when expanded near the degeneracy point takes the following
universal form as well
\beq
\label{eq:6}
H(\bk) = \pm v_F \bsigma \cdot \bk, 
\eeq
where the crystal momentum $\bk$ is measured from the degeneracy point and we will use $\hbar = c =1$ units throughout, 
except in some of the final formulas. 
$\bsigma$ is a triplet of Pauli matrices, which act in the reduced Hilbert space of the two touching bands and the sign in front 
is the topological charge of the node. 
This Hamiltonian is identical, up to a replacement of $v_F$ by the speed of light $c$, to the Hamiltonian for massless relativistic 
fermions, proposed by Hermann Weyl in 1929. 
The topological charge of the band touching node is identical to the chirality of Weyl fermions, right-handed (R) corresponding 
to the topological charge $C = + 1$ and left-handed (L) to $C = -1$. 

Eq.~\eqref{eq:6} betrays an important property of topological metals: their band eigenstates must be nondegenerate. 
This requires either broken time-reversal symmetry (a magnetic material) or a broken spatial inversion symmetry (the crystal structure lacks a center of inversion). Otherwise, all bands are doubly degenerate due to the Kramers theorem. 
On the other hand, if either time-reversal or inversion are indeed broken, Eq.~\eqref{eq:6} also makes it clear that no other 
conditions are necessary for the existence of the Weyl band touching nodes in a 3D material, since the three components of the 
crystal momentum provide the necessary three real parameters needed to make two band eigenstates coincide in energy. 
Thus Weyl nodes will occur generically in any 3D magnetic or noncentrosymmetric material, of which there are plenty. 

In this article we will review the exciting and rapidly developing subject of Weyl semimetals, focusing in response and transport 
phenomena. 
As excellent reviews of the experiments already exist, see for example Refs.~\cite{Felser_ARCMP,Hasan_ARCMP}, we will instead focus on general principles, presented using simple idealized model systems. 

\section{WEYL SEMIMETAL}
\label{sec:2}
The presence of Weyl nodes in the electronic structure is by itself of little interest. 
The important question is whether their presence leads to observable consequences. 
It is clear that such consequences will only exist when the nodes are close to the Fermi energy (Weyl metal), ideally right 
at the Fermi energy (Weyl semimetal). In addition, there should not be any other states at the Fermi energy, which may 
obscure the effects due to the Weyl nodes. 
While the presence of Weyl nodes somewhere in the band structure is guaranteed simply by broken time-reversal or 
inversion symmetries and by the three dimensionality, their location relative to the Fermi energy is not generally guaranteed by anything. 

\subsection{WEYL SEMIMETAL NEAR THE QUANTUM HALL PLATEAU TRANSITION}
\label{sec:2.1}
For a given electron density, Pauli principle, which states that two electrons can not occupy the same state, and the fact that every band has exactly the same number of states, equal to the number of unit cells in the crystal, fix the volume in the crystal momentum space, enclosed by the Fermi surface, modulo BZ volume. 
This statement is known as the Luttinger's theorem. 
Since the shape of the Fermi surface depends on the details of the band structure, it follows that the location of the Fermi energy is universally determined by the electron density and independent of other details only in insulators. 
This suggests, at least in theory, a route for finding a material with Weyl nodes at the Fermi energy: one needs to start from an insulator, which has the required broken time-reversal or inversion symmetry, and close its gap, i.e. tune the insulator to a quantum phase transition. 
The Fermi level in this case will be pinned to the band touching points by the electron density. 
As we will see this is exactly the right strategy: Weyl semimetal, i.e. a material with Weyl nodes, but no other states at the 
Fermi energy, occurs generically as an intermediate phase between two topologically distinct kinds of insulators, when a direct 
transition between them is impossible~\cite{Murakami07,Burkov11-1}. 
An important implication of this is that the most convenient way to describe Weyl semimetals theoretically is 
by expanding about such critical points. 
The simplest case of this occurs when time-reversal symmetry is violated and the transition then is between an ordinary 
3D insulator and a quantum anomalous Hall insulator with a quantized Hall conductivity. 

\begin{figure}[t]
\vspace{-2cm}
\includegraphics[width=12cm]{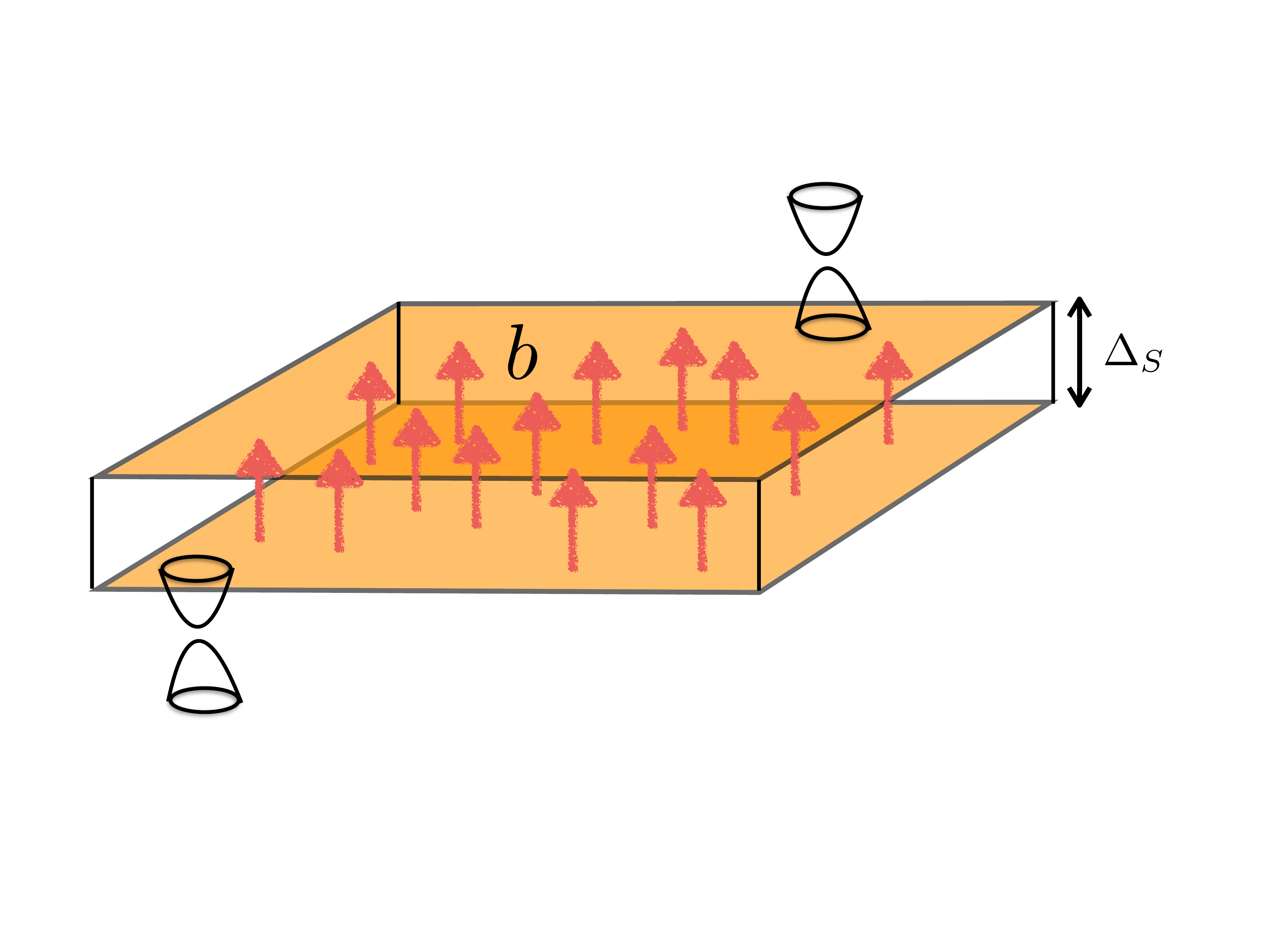}
\vspace{-2cm} 
\caption{Thin film of magnetically doped 3D TI. $\Delta_S$ is the tunneling amplitude between the top and bottom 2D Dirac 
surface states and $b$ is the spin splitting due to magnetized impurities. When $b > \Delta_S$, the film is a quantum anomalous 
Hall insulator with $\sigma_{xy} = e^2/h$. When $b < \Delta_S$, it is an ordinary insulator with zero Hall conductivity.}
\label{fig:1}
\end{figure}

A 3D quantum anomalous Hall insulator may be obtained my making a stack of 2D quantum Hall insulators~\cite{Kohmoto92}.
2D quantum anomalous Hall effect (QAHE) arises naturally in a very thin film of a 3D TI material, doped with magnetic 
impurities~\cite{Chang13,QAHE}.
A thin film of magnetically-doped 3D TI may be modeled by focusing only on the low-energy degrees of freedom, which 
are simply the 2D Dirac surface states on the top and bottom surfaces of the film, see \textbf{Figure~\ref{fig:1}}. 
The corresponding Hamiltonian reads
\beq
\label{eq:7}
H = v_F \tau^z (\hat z \times \bsigma) \cdot \bk + \Delta_S \tau^x + b \sigma^z. 
\eeq
Here $\bsigma$ refers to the electron spin, the eigenvalues of $\tau^z$ refer to the top or bottom surface degree of freedom, 
$\hat z$ is the direction, perpendicular to the plane of the film, $\Delta_S$ is the probability amplitude for tunneling between the 
top and bottom surfaces of the film, and $b$ is the exchange spin-splitting, which arises due to the presence of 
magnetized impurities. 
A simple similarity transformation $\sigma^{\pm} \rightarrow \tau^z \sigma^{\pm}$ and 
$\tau^{\pm} \rightarrow \sigma^z \tau^{\pm}$, brings this to the form
\beq
\label{eq:8}
H = v_F (\hat z \times \bsigma) \cdot \bk + (\Delta_S \tau^x + b) \sigma^z, 
\eeq
which may be further brought to a block-diagonal form by diagonalizing the $\Delta_S \tau^x$ matrix
\beq
\label{eq:9}
H_r = v_F (\hat z \times \bsigma) \cdot \bk + (b + r \Delta_S) \sigma^z, 
\eeq
where $r = \pm$. 
Each of the $2 \times 2$ blocks of Eq.~\eqref{eq:9} is the Hamiltonian of a 2D Dirac fermion with a ``mass" 
$m_r = b + r \Delta_S$. 
The 2D Dirac fermion exhibits an interesting property, which is known as the ``parity anomaly"~\cite{Redlich84,Haldane88,Ludwig94}. 
In our context this means that $H_{\pm}$ is associated with a Hall conductivity
\beq
\label{eq:10}
\sigma^r_{xy} = \frac{e^2}{2 h} \textrm{sign}(m_r),
\eeq
when the Fermi energy is in the gap between the positive and negative energy bands, obtained by diagonalizing $H_r$
\beq
\label{eq:11}
\epsilon_{r s} (\bk) = s \sqrt{v_F^2 \bk^2 + m_r^2}, 
\eeq 
with $s = \pm$.  
The ``anomaly" refers to the singular behavior of $\sigma^r_{xy}$ in the limit $m_r \rightarrow 0$. 

This implies that this system exhibits a quantum Hall plateau transition from $\sigma_{xy} = 0$ to 
$\sigma_{xy} = e^2/h$ as the ratio of $b/\Delta_s$ is varied and taking both $b$ and $\Delta_S$ to be positive for concreteness.
In other words, in 2D there exists a direct transition between a topological insulator with $\sigma_{xy} = e^2/h$ and a normal 
insulator with $\sigma_{xy} = 0$.
The critical point between the two is described by a massless 2D Dirac Hamiltonian. 
\begin{figure}[t]
\vspace{-2cm}
\includegraphics[width=12cm]{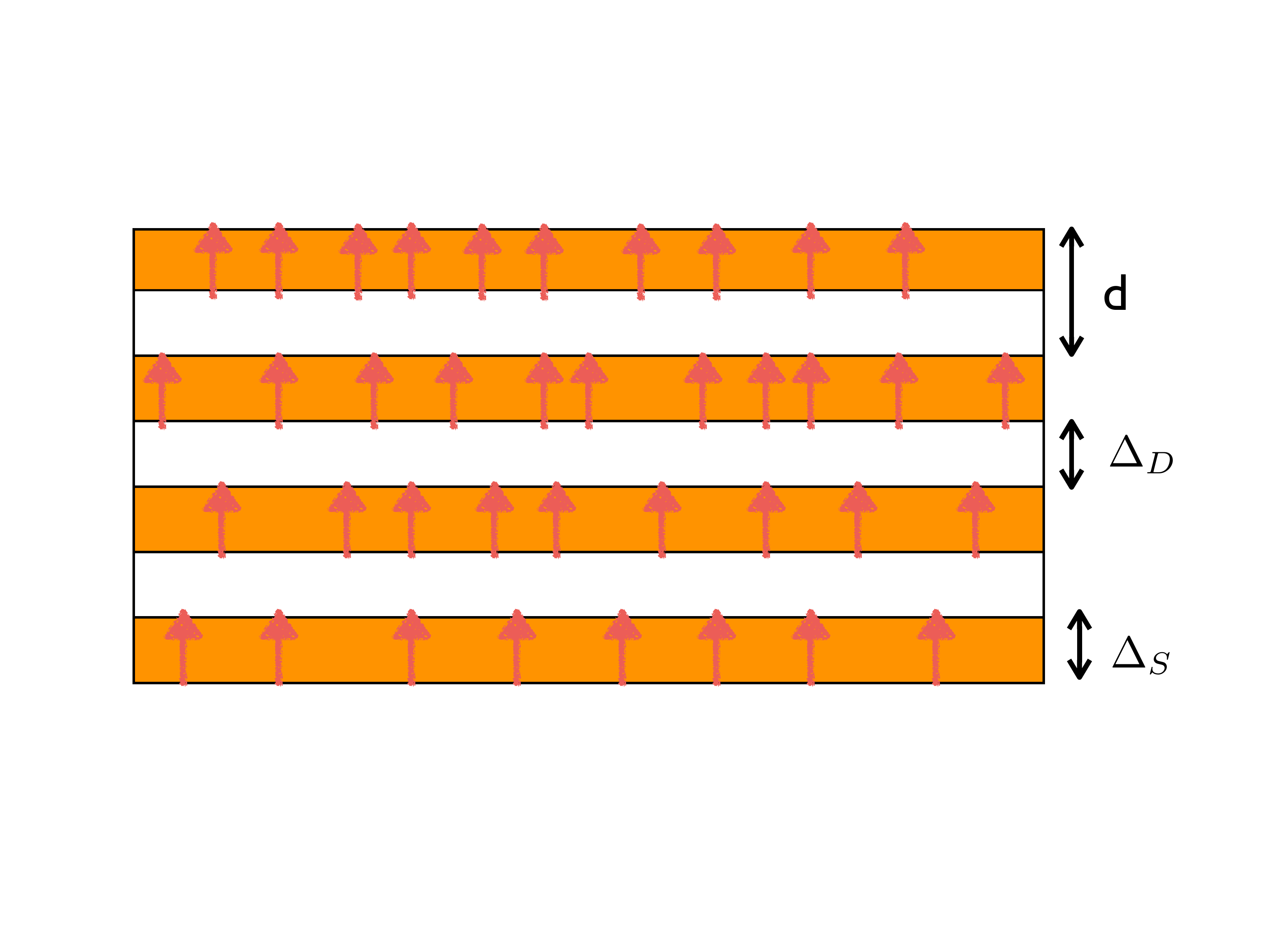}
\vspace{-2cm} 
\caption{Coupled-layer construction of an elementary Weyl semimetal. Magnetically-doped TI layers are coupled 
through insulating spacers. The tunneling amplitude between neighboring TI layers is $\Delta_D$. Further-neighbor tunneling is 
assumed to be negligibly small, leading to a particle-hole symmetric electronic structure.}
\label{fig:2}
\end{figure}

\begin{figure}[t]
\vspace{-2cm}
\includegraphics[width=12cm]{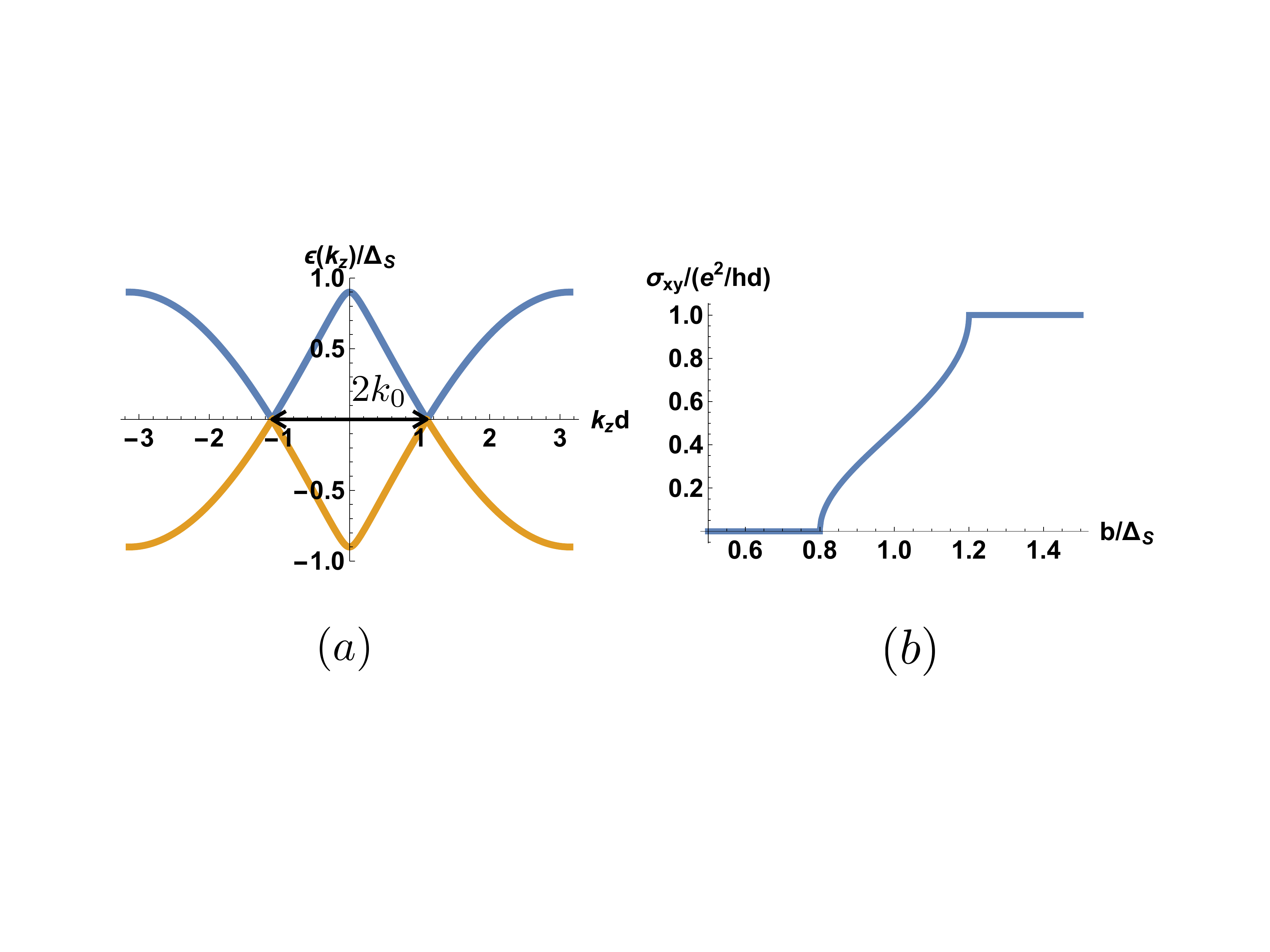}
\vspace{-2cm}
  \caption{(a) Electronic structure of the simplest Weyl semimetal, with two nodes of opposite chirality, separated by 
  a distance of $2 k_0$ along the $z$-axis in momentum space. (b) The corresponding anomalous Hall conductivity as 
  a function of $b/\Delta_S$, showing a broadened plateau transition. Weyl semimetal is an intermediate gapless phase 
  between the quantum anomalous Hall and ordinary insulators.}
  \label{fig:3}
\end{figure}

Suppose we now make a stack of such 2D layers, exhibiting QAHE, as shown in \textbf{Figure~\ref{fig:2}}. 
Let individual layers be separated by insulating 
spacers, such that the amplitude for tunneling between the adjacent surfaces of neighboring QAHE layers is $\Delta_D$, which 
we will also take to be positive.
We will take the tunneling amplitude to beyond nearest neighbor surface states to be negligibly small. 
The Hamiltonian that describes this system takes the form
\beq
\label{eq:12}
H = v_F \tau^z (\hat z \times \bsigma) \cdot \bk + [\Delta_S + \Delta_D \cos(k_z d)] \tau^x - \Delta_D \sin(k_z d) \tau^y + b \sigma^z, 
\eeq
where $d$ is the period of the resulting superlattice heterostructure in the $z$-direction. 
Making the same similarity transformation as above and partially diagonalizing the resulting Hamiltonian, we obtain
\beq
\label{eq:13}
H_r = v_F (\hat z \times \bsigma) \cdot \bk + m_r(k_z) \sigma^z, 
\eeq
where $m_r(k_z) = b + r \sqrt{\Delta_S^2 + \Delta_D^2 + 2 \Delta_S \Delta_D \cos(k_z d)} \equiv b + r \Delta(k_z)$. 
Now we see that the quantum Hall plateau transition we discussed before as a function of $b/\Delta_S$, may now happen 
``on its own" in momentum space as $k_z$ is swept through the BZ. 
Indeed $m_-(k_z)$ will change sign at $k_z^{\pm} = \pi/d \pm k_0$, where 
\beq
\label{eq:14}
k_0= \frac{1}{d} \arccos\left(\frac{\Delta_S^2 + \Delta_D^2 - b^2}{2 \Delta_S \Delta_D}\right). 
\eeq
At $\bk  = (0, 0, k^{\pm}_z)$, the two nondegenerate bands, corresponding to the eigenvalue $r = -$ touch each other, 
i.e. these are locations of two Weyl nodes, see Figure~\ref{fig:3}. 

The nodes exist as long the spin splitting $b$ is in the interval between two critical values 
$b_{c1} < b < b_{c2}$, where $b_{c1} = |\Delta_S - \Delta_D|$ and $b_{c2} = \Delta_S + \Delta_D$. 
When $b < b_{c1}$ the system is an ordinary insulator with $\sigma_{xy} = 0$, while when $b > b_{c2}$ it is a 3D quantum anomalous Hall insulator with $\sigma_{xy} = e^2/h d$. 
\begin{figure}[t]
\includegraphics[width=12cm]{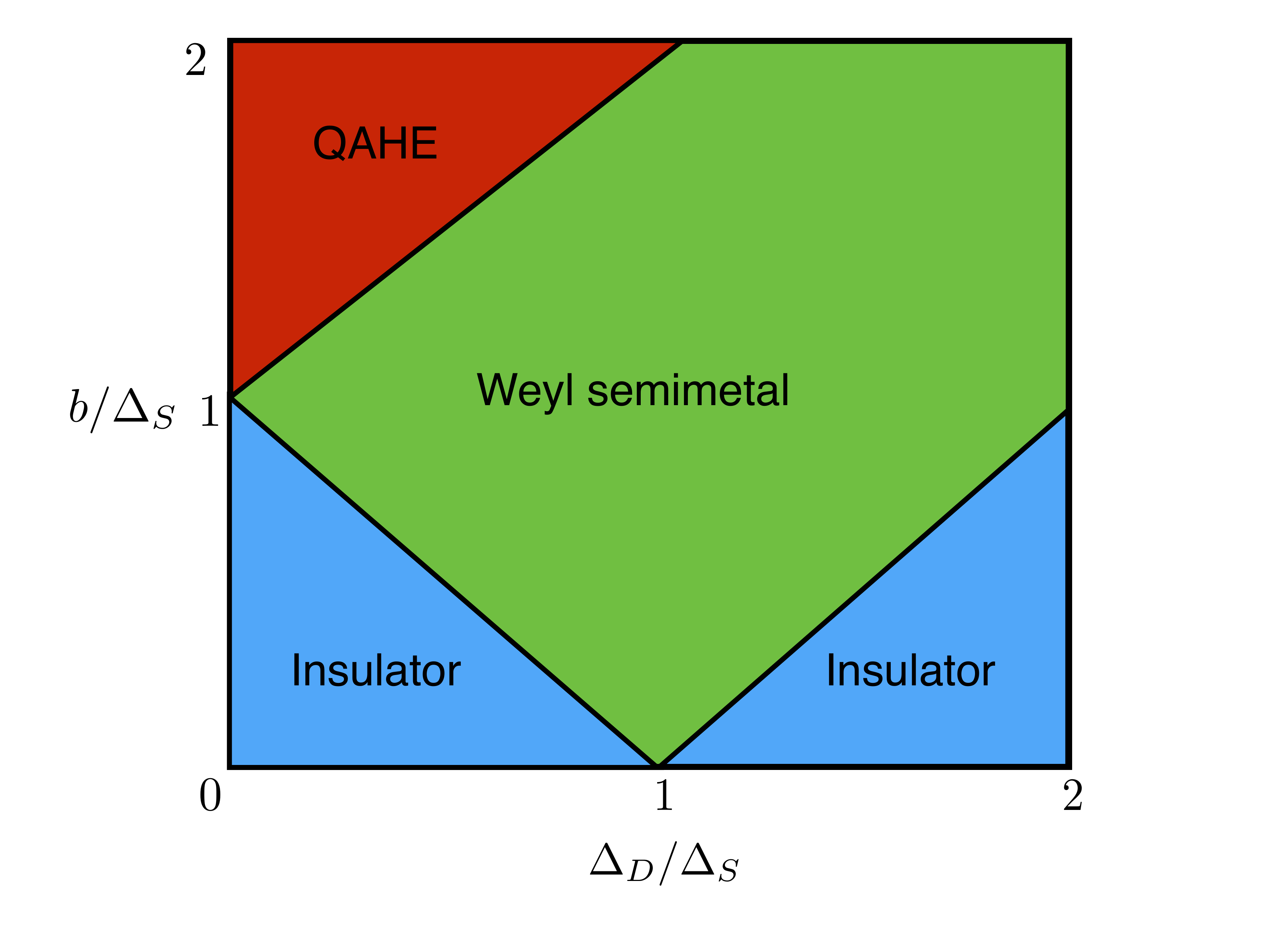}
  \caption{Phase diagram of the magnetically-doped multilayer structure. The width of the Weyl semimetal phase 
 as a function of $b/\Delta_S$ vanishes in the limit of decoupled 2D layers, in which case a direct transition between an ordinary and a quantum anomalous 
 Hall insulator exists. The maximal width is achieved once $\Delta_D/\Delta_S \geq 1$.}
  \label{fig:4}
\end{figure}
In between the heterostructure is in the intermediate Weyl semimetal phase with 
\beq
\label{eq:14.1}
\sigma_{xy} = e^2 k_0/\pi h,
\eeq
which depends only on the distance between the Weyl nodes in momentum space and varies continuously between 
$0$ and $e^2/hd$, see \textbf{Figure~\ref{fig:4}}. 
Thus, unlike in 2D, in three dimensions a direct transition between a topological insulator with nonzero quantized Hall conductivity 
and a normal insulator with zero Hall conductivity does not exist. 
The transition instead proceeds through an intermediate gapless Weyl semimetal phase. 
The system, described above, constitutes the simplest potential realization, the ``hydrogen atom" of 
Weyl semimetals. A lot of the general physical properties of Weyl semimetals may be understood by studying this 
system. 

\subsection{WEYL SEMIMETAL FROM A 3D DIRAC SEMIMETAL}
\label{sec:2.1}
A distinct, and a very useful viewpoint on Weyl semimetals is obtained in the limit when the range of the Weyl semimetal phase in 
the coupled-layer construction, described above, is maximized. 
This happens when $\Delta_D = \Delta_S$, i.e. when neighboring layers are coupled maximally strongly. 
In this limit, when time reversal symmetry is unbroken, i.e. when $b = 0$, Eq.~\eqref{eq:12} describes a Dirac semimetal, 
i.e. a state in which doubly-degenerate conduction and valence bands touch at the edge of the BZ $\bk = (0,0,\pi/d)$. 
This point is in fact a critical point between a time reversal invariant topological and normal insulator in 3D. 
Close to this critical point, when the gap at $\bk = (0,0,\pi/d)$, $|\Delta_S - \Delta_D|$, is small, Weyl semimetal phase arises most easily, as long as $b > |\Delta_S - \Delta_D|$. 
The value of this limit from the theoretical viewpoint is universality and extra symmetry (essentially Lorentz invariance) 
that emerges. 
We will discuss physics of Weyl semimetals by expanding about this ideal limit.  

Note, that this picture does not rely on any particular model of Weyl semimetal, like the superlattice model introduced above.
Indeed, let us consider a hypothetical material with four degrees of freedom per unit cell: two orbital and two spin. 
This is the minimal number of degrees of freedom necessary to obtain a pair of doubly degenerate bands,
which may exhibit a gap-closing quantum phase transition. 
Let us introduce two sets of Pauli matrices $\tau^i$ and $\sigma^i$, which will represent operators acting on the orbital and 
spin degrees of freedom correspondingly. 
In the presence of inversion symmetry, the two orbital states may always be chosen to be related to each other 
by the parity operator $P$. 
We thus take the orbital states to be the eigenstates of $\tau^z$, in which case the parity operator $P = \tau^x$. 

The most general time-reversal and parity-invariant momentum-space Hamiltonian, describing the above system, 
may be written as~\cite{Fu-Kane}
\beq
\label{eq:15}
H(\bk) = d_0(\bk) + \sum_{a = 1}^5 d_a(\bk) \Gamma^a, 
\eeq
where $\Gamma^a$ are the five matrices, realizing the Clifford algebra $\{\Gamma^a, \Gamma^b \} = 2 \delta^{ab}$, even under the product of parity and time reversal $P \Theta$.
The five $\Gamma$-matrices are given by
\beq
\label{eq:16}
\Gamma^1= \tau^x,\,\, \Gamma^2 = \tau^y,\,\, \Gamma^3 = \tau^z \sigma^x,\,\, \Gamma^4 = \tau^z \sigma^y,\,\,
\Gamma^5 = \tau^z \sigma^z. 
\eeq
It is clear that $\Gamma^1$ is even under both parity and time reversal separately, while $\Gamma^{2-5}$ are odd 
under both separately. 
The eigenvalues of Eq.~\eqref{eq:15} form two pairs of doubly-degenerate bands
\beq
\label{eq:17}
\epsilon_{\pm}(\bk) = d_0(\bk) \pm \sqrt{\sum_{a = 1}^5 |d_a(\bk)|^2}.
\eeq 
The two pairs of bands may be forced to touch at a crystal symmetry related set of time-reversal invariant momenta (TRIM) 
$\bGamma$. 
In this case we have $d_{2-5}(\bGamma) = 0$ automatically, while $d_1(\bGamma) \equiv m$ needs to be tuned 
to zero, if no additional symmetries are present. 
Since $d_{2-5}$ are all odd under parity, their Taylor expansion near $\bGamma$ will generically start 
from linear terms, while the expansion of $d_{0,1}(\bk)$ will start from a constant and continue with a quadratic term. 
Then the momentum space Hamiltonian Eq.~\eqref{eq:15}, expanded to leading order near TRIM $\bGamma$, takes the 
following from
\beq
\label{eq:18}
H (\bk)= \gamma^0 \gamma^i k_i + m \gamma^0, 
\eeq
where $\gamma^0 \equiv \Gamma^1$, while the other three Dirac matrices $\gamma^i$ are defined by this expansion. 
We have absorbed the velocity coefficients into the definition of the momentum components $k_i$, which are 
measured from $\bGamma$, and taken $d_0(\bGamma)$ to define the overall zero of energy. 
In the case of the superlattice model, the Dirac point occurs at $\bGamma = (0,0,\pi/d)$ when $\Delta _S = \Delta_D$ and $m = \Delta_S - \Delta_D$. 
The gamma matrices in this case are given by $\gamma^0 = \tau^x$, $\gamma^1 = i \tau^y \sigma^y$, $\gamma^2 = - i \tau^y \sigma^x$ and $\gamma^3 = i \tau^z$. 

The utility of this ``relativistic" representation of the Hamiltonian, apart from emphasizing a connection to the relativistic quantum 
mechanics and the emergent Lorentz invariance, is that we may now easily classify 
time-reversal and inversion symmetry breaking perturbations to $H$ in terms of the gamma matrices 
and their descendants and clearly see their physical meaning. 
The ten possible perturbations to $H$ are 
\beq
\label{eq:19}
\gamma^0 \gamma^{\mu} \gamma^5,\,\, \gamma^0 \sigma^{\mu \nu},
\eeq
where  $\sigma^{\mu \nu} = (i/2) [\gamma^{\mu}, \gamma^{\nu}]$, $\mu, \nu = 0,1,2,3$ and $\gamma^5 = i \gamma^0 \gamma^1 \gamma^2 \gamma^3$ is the chirality operator. 
The significance of $\gamma^5$ is that the operator 
\beq
\label{eq:20}
P = \frac{1}{2}(1 + \gamma^5), 
\eeq
projects out the two Weyl fermion components of the Dirac fermion. 
The right-handed Weyl fermion corresponds to the eigenvalue $+1$ of $\gamma^5$ while 
the left-handed Weyl fermion to the eigenvalue $-1$. 

Of the ten perturbations in Eq.~\eqref{eq:19}, there are six that are odd under time reversal symmetry but even under parity: 
$\gamma^0 \gamma^i \gamma^5$, $\gamma^0 \sigma^{ij}$; and four that are even under time reversal but odd under parity: 
$\gamma^5$, $\gamma^0 \sigma^{0 i}$. 
In other words, $\gamma^0 \gamma^i \gamma^5$ transforms as an axial current vector, $\gamma^5$ is a pseudoscalar, 
$\gamma^0 \sigma^{ij}$ transforms 
as magnetization, perpendicular to the $i j$ plane, and $\gamma^0 \sigma^{0 i}$ transforms as 
electric polarization along the $i$ axis. 
As a result, perturbation of the type $\bb \cdot \gamma^0 \bgamma \gamma^5$ acts as a chiral vector potential, shifting 
the location of Weyl fermions of different chirality to different momenta, i.e. precisely creating a Weyl semimetal. 
A $b_0 \gamma^0 \gamma^0 \gamma^5 = b_0 \gamma^5$ term acts as a time-component of the chiral gauge 
field, shifting the two Weyl components of the Dirac fermion to different energies. 
The term of the type $b_i \epsilon_{ijk} \gamma^0 \sigma^{jk}$ creates nodal lines, in the plane, perpendicular to the vector $\bb$. 
Finally, a $b_i \gamma^0 \sigma^{0i}$ term, which breaks inversion symmetry, creates Weyl nodes, which in this case come 
in multiples of four~\cite{Halasz12} (to obtain these, the inclusion of terms which are higher order in momentum and break the continuous rotational symmetry of the low energy Dirac Hamiltonian Eq.~\eqref{eq:18} down to a discrete crystal symmetry, is 
necessary). 
 
\section{TRANSPORT IN WEYL METALS}
\label{sec:3}
Perhaps the most important consequence of the nontrivial electronic structure topology is that it may lead to 
unique transport and more generally response to external probes. 
A famous example of this is the quantized Hall conductivity of a 2D quantum Hall insulator, described 
in Section~\ref{sec:2.1}. 
This quantization is a consequence of the fact that the transport in a quantum Hall insulator occurs entirely in the 
one dimensional chiral edge states. 
Since the branches of different chirality are separated to opposite edges of a macroscopic sample, no scattering 
between them is possible. 
This means that chiral charges, corresponding to the edge states, are separately conserved, which leads to universal 
quantized transport. 

As will be shown in this section, similar ideas are applicable to {\em bulk transport} in Weyl semimetals. 
This is a consequence of the (approximate) separate conservation of the number of electrons of different chirality, 
which becomes most precise in the vicinity of the 3D Dirac critical point, described in the previous section. 
These (quasi-)universal transport phenomena in Weyl metals may be regarded as being a consequence of chiral anomaly, 
a well-known and important phenomenon in particle physics~\cite{Adler69,Jackiw69}, which has now found its way into 
condensed matter. 

\subsection{CHIRAL ANOMALY}
\label{sec:3.1}
Chiral anomaly refers to nonconservation of chiral charge for massless relativistic particles, when conservation is naively 
expected based on symmetry. 
Indeed, consider the Dirac Hamiltonian Eq.~\eqref{eq:18} in the limit when the mass $m = 0$
\beq
\label{eq:21}
H (\bk)= \gamma^0 \gamma^i k_i, 
\eeq
From the anticommutation property of the individual gamma matrices, it immediately follows that the $\gamma^5$ 
matrix commutes with the Hamiltonian 
\beq
\label{eq:22}
[H, \gamma^5] = 0. 
\eeq
This expresses an exact extra conservation law that exists for massless relativistic fermions: conservation of the chiral charge. 
This conservation law implies that the chiral current, defined by $j^{\mu}_5 = \psi^\dg \gamma^0 \gamma^{\mu} \gamma^5 \psi^\pdg$, where $\psi^\dg$ is the electron creation operator,
must satisfy the continuity equation
\beq
\label{eq:23}
\partial_{\mu} j^{\mu}_5 = 0, 
\eeq
which must hold in addition to the regular electric charge conservation law
\beq
\label{eq:24}
\partial_{\mu} j^{\mu} = 0, 
\eeq
where $j^{\mu} = \psi^\dg \gamma^0 \gamma^{\mu} \psi^\pdg$. 
In reality, the seemingly obvious chiral charge conservation turns out to be violated 
in the second-quantized theory, when the negative-energy Dirac sea is filled with electrons, in the presence 
of an applied electromagnetic field
\beq
\label{eq:25}
\partial_{\mu} j^{\mu}_5 = \frac{e^2}{16 \pi^2} \epsilon^{\mu \nu \alpha \beta} F_{\mu \nu} F_{\alpha \beta} = 
\frac{e^2}{2 \pi^2} {\bf E} \cdot {\bf B}. 
\eeq

The simplest way to understand this is to consider a massless Dirac fermion, described by  Eq.~\eqref{eq:21}, in the 
presence of a constant magnetic field, applied in, say, $z$-direction.
\begin{figure}[t]
\includegraphics[width=12cm]{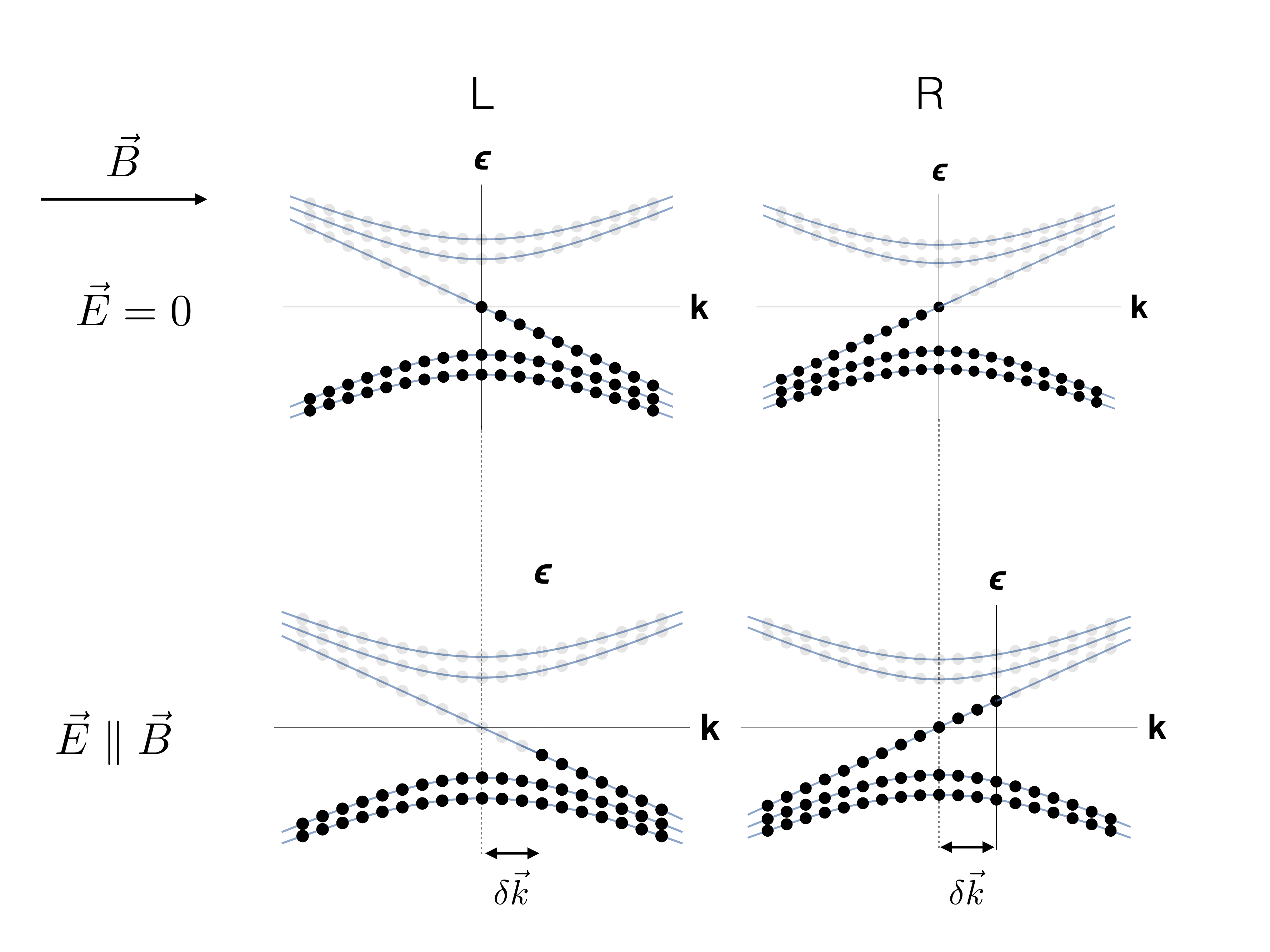}
  \caption{Illustration of the chiral anomaly, based on Landau level spectrum of Weyl fermions in an external magnetic field. Top panel: Energy spectrum of the left-handed (L) and the right-handed (R) fermions in equilibrium in the presence of a magnetic field $\vec B$. Filled states with negative energy are shown by black dots, while empty states with positive energy by gray dots. Bottom panel: Same spectrum, but in the presence, in addition, of an electric field $\vec E$, parallel to the magnetic field $\vec B$. 
 All states have been displaced in momentum space by an amount $\delta \vec k \sim - \vec E$ from their equilibrium locations. 
 As a consequence, right-handed particles and left-handed antiparticles have been produced.}
  \label{fig:5}
\end{figure}
The resulting Landau level structure for the two independent Weyl fermion components is shown in Figure.~\ref{fig:5}. 
Most importantly, the chirality of the Weyl fermions is reflected in their Landau level spectrum, manifesting in the 
presence of a special Landau level in each case, whose dispersion is given by
\beq
\label{eq:26}
\epsilon(k_z) = \pm k_z, 
\eeq 
where the sign in front is minus the chirality (eigenvalue of the $\gamma^5$ matrix) of the corresponding 
Weyl fermion. 
Suppose we now apply an electric field ${\bf E}$ in the same direction as the magnetic field. 
The electrons will be accelerated at a rate $e E$, in the $\hat E$ direction (we will imagine electrons to be positively 
charged for the sake of this argument, just to avoid some awkward minus signs, the final result is independent 
of the sign of the charge) which, due to the presence of the chiral 
Landau levels, results in the following rates of the right- and left-handed charge generation
\beq
\label{eq:27}
\frac{\partial n_{R, L}}{\partial t} = \pm \frac{e^2}{4 \pi^2} {\bf E} \cdot {\bf B}, 
\eeq
where the factor of $e B /4 \pi^2$ arises from the density of states of the chiral Landau levels.  
This precisely corresponds to Eq.~\eqref{eq:25}, with $j^0_5 \equiv n_5 = n_R - n_L$. 

This may seem fairly inconsequential as the chiral charge by itself is not a directly observable quantity. 
In order to see that the chiral anomaly does in fact have observable consequences, it is useful 
to note that Eq.~\eqref{eq:25} may be obtained from the following action for the electromagnetic field
\beq
\label{eq:28}
S = - \frac{e^2}{4 \pi^2} \int d t \, d^3r \, b_{\mu} \epsilon^{\mu \nu \alpha \beta} A_{\nu} \partial_{\alpha} A_{\beta}.
\eeq
Here $b_{\mu}$ are components of the ``chiral gauge field", which couples linearly to the chiral current as $b_{\mu} j^{\mu}_5$, just 
as ordinary gauge field couples linearly to the ordinary electric current as $A_{\mu} j^{\mu}$. 
Eq.~\eqref{eq:25} is obtained by calculating the functional derivative of $S$ with respect to the chiral 
gauge field $b_{\mu}$, which gives the chiral current $j^{\mu}_5$, and taking the divergence. 

\begin{figure}[t]
\vspace{-2cm}
\includegraphics[width=12cm]{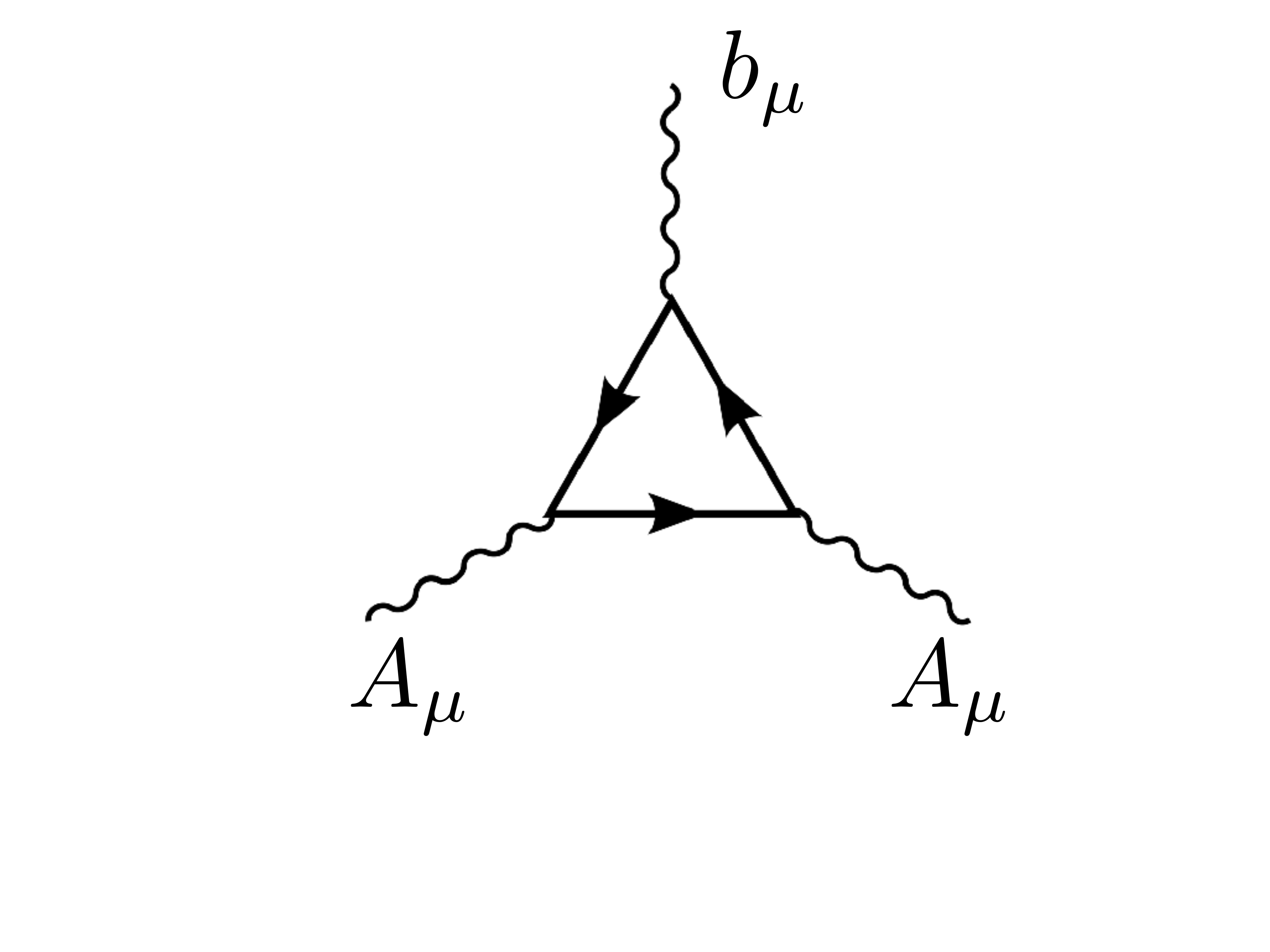}
\vspace{-2cm}
  \caption{Graphical representation of Eq.~\eqref{eq:28}, known as the triangle diagram. The three vertices correspond 
  to the chiral current (top vertex) and two ordinary electrical currents (bottom vertices).}
  \label{fig:6}
\end{figure}

Eq.~\eqref{eq:28} corresponds to the famous ``triangle diagram", shown in Figure~\ref{fig:6}, which is a diagram 
with three vertices, two corresponding to the ordinary charge current $j^{\mu}$ and one to the chiral current $j^{\mu}_5$. 
The original discovery of the chiral anomaly by Adler~\cite{Adler69} and by Bell and Jackiw~\cite{Jackiw69} in 1969 
was made when evaluating this diagram in the context of the problem of neutral pion decay into two 
photons
\beq
\label{eq:29}
\pi^0 \rightarrow 2 \gamma.  
\eeq
This process, which naively should be very slow due to small masses of the quarks (pion is a quark-antiquark pair), in fact 
is very fast, leading to the neutral pion lifetime being many orders of magnitude smaller that that of the two 
charged pions. 
The explanation for this turned out to be the chiral anomaly, which now, almost 50 years later, has found its way into 
condensed matter physics as well. 

The importance of Eq.~\eqref{eq:28} stems from the fact that it clearly has a topological origin, as it contains gauge fields, coupled 
through the fully antisymmetric tensor, and multiplied by a universal coefficient. 
One may thus expect the corresponding response to be robust and detail-independent, which is always of great interest, and 
of potential technological importance. 
In the following section we will describe these universal electromagnetic response phenomena, resulting from the chiral 
anomaly, as applicable to Weyl metals. 
\subsection{CHIRAL ANOMALY AND TRANSPORT IN WEYL METALS}
\label{sec:3.2}
To see the observable consequences of the chiral anomaly, we may vary the action $S$ with respect to the electromagnetic 
gauge fields $A_{\mu}$. 
This gives the following expression for the chiral anomaly contribution to the measurable charge current
\beq
\label{eq:30}
j^{\nu} = \frac{e^2}{2 \pi^2} b_{\mu} \epsilon^{\mu \nu \alpha \beta} \partial_{\alpha} A_{\beta}. 
\eeq
Suppose the chiral gauge field $b_{\mu}$ has both spatial and temporal components $b_{\mu} = (b_0, -\bb)$. 
Then we obtain
\beq
\label{eq:31}
\bj = \frac{e^2}{2 \pi^2} (\bE \times \bb), 
\eeq
and 
\beq
\label{eq:32}
{\bf j} = \frac{e^2}{2 \pi^2} b_0 {\bf B}. 
\eeq
To understand the meaning of Eqs.~\eqref{eq:31} and Eq.~\eqref{eq:32}, we first need to understand the physical 
meaning if the chiral gauge field $b_{\mu}$ in the condensed matter context. 
Suppose we add a term $b_{\mu} j^{\mu}_5$ to the Dirac Hamiltonian Eq.~\eqref{eq:18}
\beq
\label{eq:33}
H(\bk) = \gamma^0 \gamma^i (k_i - b_i \gamma^5) + m \gamma^0 + b_0 \gamma^5. 
\eeq
It is then clear that, when $m = 0$, the chiral gauge field simply shifts the two Weyl fermion components 
of the massless Dirac fermion to different locations in momentum space ($\bb$) or in energy ($b_0$). 
In particular, a nonzero $\bb$ creates a Weyl semimetal, in which two Weyl fermions are separated by $2 |\bb|$ in 
momentum space along the direction of the vector $\bb$ in momentum space. 
Chiral anomaly in this context means that, while naively the separation between the Weyl nodes may be eliminated 
by a chiral gauge transformation
\beq
\label{eq:34}
\psi \rightarrow e^{i \gamma^5 \theta}  \psi, \,\, b_{\mu} \rightarrow b_{\mu} + \partial_{\mu} \theta, 
\eeq
where $\partial_{\mu} \theta = - b_{\mu}$, in reality this is impossible: the chiral transformation 
leaves behind precisely the topological term of Eq.~\eqref{eq:28}. 
The mass term $m \gamma^0$ violates the chiral gauge invariance explicitly.
However, gapless Weyl points still exist as long as $|\bb| > m$, separated in momentum space by $2 \sqrt{\bb^2 - m^2}$. 
Chiral gauge invariance is then effectively restored in the limit $|\bb| \gg m$. 

An observable consequence of the momentum-space separation between the Weyl nodes is the topological 
current, given by Eq.~\eqref{eq:31}. This describes anomalous Hall effect (AHE) with the transverse conductivity given by
\beq
\label{eq:35}
\sigma_{xy} = \frac{e^2}{h} \frac{2 |\bb|}{2 \pi}, 
\eeq
which is exactly Eq.~\eqref{eq:14.1} we have obtained before based on analogy with the quantum Hall plateau transition. 
We thus see that AHE in magnetic Weyl semimetals may be regarded as being a consequence of chiral anomaly. 

This is not simply a rederivation of Eq.~\eqref{eq:14.1}, however, as Eq.~\eqref{eq:35}, being a consequence of the chiral 
anomaly, implies significantly more universality of this result, than may naively be expected. 
Indeed, a quantized value of the transverse conductivity of a quantum Hall insulator is protected by the spectral gap. 
As long as the Fermi energy is within the gap (which may be a mobility gap only), the quantized value of the Hall conductivity 
can not change. 
This is not the case in a gapless system, such as a Weyl semimetal. 
Uncontrollable impurities and other imperfections will always shift the location of the Fermi energy, even if it exactly 
coincides with the Weyl nodes in a perfect stoichiometric sample. 
In general, we may thus expect bulk states to contribute to the Hall conductivity and the universal (in the sense of being 
dependent only on the Weyl node separation) result of Eqs.~\eqref{eq:14.1}, \eqref{eq:35} will not hold. 
It turns out, however, that Eq.~\eqref{eq:35} is in fact much more robust than would normally be expected, which may 
be thought of as a consequence of the nonrenormalization of the chiral anomaly. 

The universal topological result for the action, describing the electromagnetic response, Eq.~\eqref{eq:28}, holds as 
long as the field $b_{\mu}$, which couples to the chiral current as $b_{\mu} j^{\mu}_5$, enters the Hamiltonian as a 
chiral gauge field, with nothing violating chiral gauge invariance, except the chiral anomaly itself. 
Anything that explicitly violates chiral gauge invariance, i.e. the mass term $m \gamma^0$, or any term, containing higher 
powers of momentum in the Hamiltonian Eq.~\eqref{eq:33}, will produce nonuniversal corrections to Eq.~\eqref{eq:28}. 
The reverse is true as well: anything that does not violate chiral gauge invariance, e.g. scalar (i.e. independent of the 
orbital and spin indices on which the gamma matrices are acting) impurity scattering potential, electron-electron interactions, 
deviation of the electron density from stoichiometry, etc., will not change the action, and the topological currents, obtained from it. 
This is somewhat similar to the universality of the Luttinger liquid physics in one-dimensional systems, which may also be thought as consequence of the chiral anomaly~\cite{Cheianov98}. 

Let us now shift our attention to the second kind of topological current that follows from the chiral anomaly, given by 
Eq.~\eqref{eq:32}. 
This equation describes a current, driven by an applied magnetic field and the chiral scalar potential $b_0$, which 
shifts the energies of the Weyl fermions in opposite directions depending on their chirality. 
This is quite troubling, since such an energy difference may exist even in equilibrium, if, for example, both time-reversal 
and inversion symmetry are violated~\cite{Zyuzin12-2}. 
Eq.~\eqref{eq:32} would then seem to imply an equilibrium current, driven by an applied magnetic field, which
certainly is not possible~\cite{Franz13}. 
In fact, such an equilibrium current would be possible, if the Weyl fermions were realized as edge states on a 3D boundary of 
a 4D quantum Hall insulator~\cite{Zhang01}. 
In this case the edge states of a 4D quantum Hall insulator slab are precisely the two 3D Weyl fermions of opposite chirality, 
existing on the ``top" and ``bottom" surfaces of the slab. 
Applying a voltage $V$ between the two surfaces and a constant magnetic field parallel to the surface, would lead precisely 
to an equilibrium current given by
\beq
\label{eq:36}
{\bf j} = \frac{e^2}{2 \pi^2} e V \, {\bf B}, 
\eeq
which is the 4D analog of the 2D quantized Hall current (also an equilibrium current). 
What makes an equilibrium current possible in both cases is the spatial separation of the 1D or 3D chiral fermions 
to opposite sides of a macroscopic sample. 
This makes it possible for the two chiral fermions to be in equilibrium independently from each other and at distinct chemical 
potentials, which leads to an equilibrium current. 

In a bulk 3D Weyl metal, however, the Weyl fermions are separated in momentum rather than real space and 
can not be in equilibrium at different chemical potentials independently from each other. 
Thus Eq.~\eqref{eq:32} may only describe the low-frequency limit of a dynamical nonequilibrium response~\cite{Chen13}. 
This is best handled within the framework of the density response theory. 
\subsection{Anomalous density response in Weyl metals}
\label{sec:3.3}
We will start this section by reminding the reader fundamentals of the density response in a metal. 
While the response at short length and time scales is highly complicated and nonuniversal, it becomes 
very simple and universal on large length and time scales, much longer than the scattering time $\tau$ and the mean free 
path $\ell = v_F \tau$. 
Only conserved quantities are important in this limit and their response is governed by simple hydrodynamic equations, which involve only the leading order temporal and spatial derivatives. 
The only conserved quantity in a typical metal is the charge density, which obeys the diffusion equation
\beq
\label{eq:37}
\frac{\partial n}{\partial t} = D \boldsymbol \nabla^2 (n + g V). 
\eeq
Here $n$ is the electron density, $D \sim v_F \ell$ is the diffusion coefficient, $g$ is the density of states at the Fermi energy 
and $V$ is the external electrostatic potential. 
The two terms on the right hand side of the diffusion equation correspond to two contribution to the electric current, diffusion 
and drift
\beq
\label{eq:38}
\bj = e D \bnabla n + e g D \bnabla V = e D \bnabla n + \sigma \bE, 
\eeq
where we have taken account of the Einstein relation $\sigma = e^2 g D$, which connects conductivity and the diffusion coefficient. 
Fourier transforming Eq.~\eqref{eq:38}, we obtain the following expression for the density response function
\beq
\label{eq:39}
\chi(\bq, \omega) = \frac{n(\bq, \omega)}{V(\bq, \omega)} = - g \frac{D \bq^2}{D \bq^2 - i \omega}. 
\eeq
A characteristic feature of this response function is the ``diffusion pole" at $i \omega = D \bq^2$, which arises fundamentally 
from the particle number conservation. 
An important property of the response function, that can be seen from Eq.~\eqref{eq:39}, is that $\chi(\bq, 0) = - g$. 
$\chi(\bq, 0)$ describes equilibrium redistribution of the charge density in response to an applied inhomogeneous external potential. 
This means that the combination $n + g V$, that appears on the right hand side of the diffusion equation~\eqref{eq:37}, 
has the meaning of a nonequilibrium part of the charge density. 

Weyl metals are distinguished from ordinary metals by the emergence of an extra conserved quantity, the chiral charge, which 
in the simplest case of two Weyl nodes may be thought of the as the difference between the densities of the right-handed 
and left-handed fermions. 
Chiral anomaly also implies that in the presence of an external magnetic field the chiral and the ordinary electric charge 
will be coupled. 
The density response function is then a $2 \times 2$ matrix~\cite{Burkov_lmr_prl,Burkov_lmr_prb} and the corresponding 
transport equations read
\beqa
\label{eq:40}
\frac{\partial n}{\partial t}&=&D \bnabla^2 (n + g V) + \Gamma \bB \cdot \bnabla (n_c + g V_c), \nonumber \\
\frac{\partial n_c}{\partial t}&=&D \bnabla^2 (n_c + g V_c) - \frac{n_c + g V_c}{\tau_c} + \Gamma \bB \cdot \bnabla (n + g V). 
\eeqa
Here $\Gamma = e/2 \pi^2 g$ is a new transport coefficient, characterizing Weyl metals, which describes the coupling 
between the chiral and the electric charge, induced by the chiral anomaly. 
Note that 
\beq
\label{eq:41}
e g \Gamma \bB \cdot \bnabla V = \frac{e^2}{2 \pi^2} \bE \cdot \bB, 
\eeq
which is identical to the right hand side of Eq.~\eqref{eq:25}. 
This means that the chiral anomaly is nonrenormalized even in the diffusive transport regime at finite electron density 
and in the presence of impurity scattering. 
$\tau_c$ is the relaxation time for the chiral charge. The presence of a relaxation term in Eq.~\eqref{eq:40} 
reflects the fact that the chiral charge in a Weyl metal is never exactly conserved. 
Its approximate conservation is an emergent low-energy property, which holds when the Fermi energy is close 
to the band-touching nodes. 
In this regime the chiral relaxation time is very long. For the multilayer model, described in Section~\ref{sec:2}, 
it is given by
\beq
\label{eq:42}
\frac{1}{\tau_c} \sim \frac{\epsilon_F^2}{\Delta_S^2 \tau}, 
\eeq
where $\epsilon_F$ is the Fermi energy. In general, $\Delta_S$ is a parameter of the order of the bandwidth. 
Note that the smallness of the chiral relaxation rate is not a consequence of separation between the Weyl 
fermions in momentum space, but rather of orthogonality of their wavefunctions (in a sense, separation in Hilbert space), 
which arises due to the emergent conservation of chirality. 
Finally, $V_c$ is the ``chiral electrostatic potential", i.e. an external potential, which couples linearly to the chiral charge. 

Since the electric charge is strictly conserved, the first of Eq.~\eqref{eq:40} must have the form of a continuity 
equation. This immediately implies the following expression for the electric current in a Weyl metal
\beq
\label{eq:43}
\bj = e D \bnabla (n + g V) + e \Gamma (n_c + g V_c) \bB. 
\eeq
The first term in Eq.~\eqref{eq:43} represents the standard diffusion and drift contributions to the electric current. 
The second term is due to the chiral anomaly and represent a contribution to the electric current, proportional to the 
nonequilibrium chiral charge density $n_c + g V_c$ and to the applied magnetic field
\beq
\label{eq:44}
\bj_{CME} = e \Gamma (n_c + g V_c) \bB. 
\eeq
This arises entirely from the chiral lowest Landau level and is known as the Chiral Magnetic Effect (CME)~\cite{Kharzeev08,Son12}. 
This is precisely equivalent to the second topological current, arising from the chiral anomaly, given by Eq.~\eqref{eq:33}, if 
we identify $b_0 = (n_c + g V_c)/g$. 
Unlike Eq.~\eqref{eq:33} however, which naively could be interpreted as predicting an equilibrium current, Eq.~\eqref{eq:44}
explicitly contains only nonequilibrium part of the chiral charge density. 

An important property of the coupled transport equations~\eqref{eq:40} is that they contain several distinct length scales, 
which leads to complex 
non-Ohmic scale dependence of the sample conductance. 
This complex scale dependence is a unique feature of Weyl metals and is a macroscopic manifestation 
of the chiral anomaly. 
The first fundamental length scale in a Weyl metal is a magnetic field induced length scale
\beq
\label{eq:45}
L_m = \frac{D}{\Gamma B}. 
\eeq
Note that this is distinct from the magnetic length $\ell_B = 1/\sqrt{e B}$. 
The physical meaning of this new field-induced length scale will become clear below. 
The second important length scale is the chiral charge diffusion length
\beq
\label{eq:48}
L_c = \sqrt{D \tau_c}. 
\eeq

To understand the role the two length scales play in transport, we solve Eq.~\eqref{eq:40}, assuming a cubic sample of volume $L^3$ and taking magnetic field to be aligned with the current for simplicity. 
One obtains the following expression for the scale-dependent conductance~\cite{Altland15}
\beq
\label{eq:49}
G(L) = \frac{e^2 N_{\phi}}{2 \pi} f\left(\frac{L}{L_m}, \frac{L}{L_c}\right), 
\eeq
where $N_{\phi} = L^2/2 \pi \ell_B^2$ is the number of flux quanta, penetrating a cross section of the sample, perpendicular to the direction of the magnetic field, which is also the same as the number of states in the lowest Landau level. 
The scaling function $f(x,y)$ is given by
\beq
\label{eq:50}
f(x,y) = \frac{(1+y^2/x^2)^{3/2}}{\frac{y^2}{2 x} \sqrt{1 + y^2/x^2} + \textrm{tanh}\left(\frac{x}{2}\sqrt{1+y^2/x^2}\right)}. 
\eeq
When $L_m \gg L_c$, or equivalently $x \ll y$, the magnetic field has only a negligible effect on the conductance and 
the scaling function has simple asymptotics 
\beq
\label{eq:51}
f(x, y) \approx \frac{2}{x}
\eeq 
Eq.~\eqref{eq:49} then gives the usual Ohmic conductance
\beq
\label{eq:52}
G(L) = e^2 g D L.
\eeq
Magnetoconductance is strong in the opposite limit when $L_m \ll L_c$ (however, we still assume quasiclassical 
condition $k_F \ell_B \gg 1$ holds), or when $x \gg y$. 
In this regime the scaling function simplifies to
\beq
\label{eq:53}
f(x,y) \approx \frac{1}{\frac{y^2}{2 x} + \textrm{tanh}(x/2)}. 
\eeq
For $L < L_c$ this exhibits a crossover from diffusive $L < L_m$ regime with regular Ohmic conductance given by
Eq.~\eqref{eq:52}, to quasiballistic conductance when $L > L_m$
\beq
\label{eq:54}
G(L) = \frac{e^2 N_{\phi}}{2 \pi}, 
\eeq
which is the conductance of an effective one-dimensional system with $N_{\phi}$ conduction channels. 
Physically, this corresponds to a regime in which the conductance is dominated by the chiral lowest Landau level. 
Upon further increase of the sample size, a crossover back into the diffusive regime happens when $y^2 > x$, or 
equivalently when the sample size exceeds yet another distinct length scale
\beq
\label{eq:55} 
L > L_* = \frac{L_c^2}{L_m} \gg L_c. 
\eeq
The new length scale $L_*$ may be obtained from equating the number of states in a sample of volume $L_*^3$
in the energy interval $1/\tau_c$, which is given by $g L_*^3 /\tau_c$, to the number of states in the lowest Landau 
level $N_{\phi}$. 
In this regime the conductance is dominated by the field-dependent chiral anomaly contribution, but with $\propto B^2$, instead 
of $\propto B$, dependence
\beq
\label{eq:56}
G(L) = \frac{e^4 \tau_c B^2}{4 \pi^4 g} L \equiv \chi B^2 L. 
\eeq
When $L_m > L_c$ this result still holds in the infinite sample limit, but with Eq.~\eqref{eq:56} only representing 
a subdominant correction to the Ohmic conductance. 
It is quite remarkable that Eq.~\eqref{eq:40} gives rise to such a rich and nontrivial scaling behavior of the conductance 
with three distinct length scales!

Generalizing Eq.~\eqref{eq:56} to an arbitrary field direction we obtain the expression for the longitudinal magnetoconductivity
that was obtained originally implicitly in the limit of an infinite sample size~\cite{Spivak12,Burkov_lmr_prl,Burkov_lmr_prb}
\beq
\label{eq:57}
\Delta \sigma_{xx} = \chi B^2 \cos^2 \theta, 
\eeq
where $\theta$ is the angle between the current and the magnetic field. 
Similar phenomena are also expected theoretically and observed in thermoelectric transport~\cite{Fiete14,GdPtBi,Andreev16,Sachdev16}.

In fact, things are a bit more complex than Eq.~\eqref{eq:57} predicts in a finite sample with boundaries~\cite{Burkov16}. 
Indeed, suppose current is fed into the sample in the $x$-direction and the magnetic field is rotated in the $xy$-plane. 
Then, assuming an infinite sample in the $x$-direction, we have
\beqa
\label{eq:58}
&&\sigma E_x + \chi (E_x B_x + E_y B_y) B_x = j_x, \nonumber \\
&&\sigma E_y + \chi (E_x B_x + E_y B_y) B_y = 0. 
\eeqa
This implies, that once the magnetic field is rotated away from the $x$-direction, i.e. from perfect alignment with the current, 
chiral anomaly leads to the development of a transverse (with respect to the direction of the current) electric field
\beq
\label{eq:59}
E_y = - \frac{\chi B_x B_y}{\sigma + \chi B_y^2} E_x, 
\eeq
This, in turn, affects longitudinal conductivity, decreasing it compared to Eq.~\eqref{eq:57}. 
Substituting Eq.~\eqref{eq:59} into the first of Eq.~\eqref{eq:58}, we obtain~\cite{Burkov16}
\beq
\label{eq:60}
\Delta \rho^{-1}_{xx} = \frac{\sigma \chi B_x^2}{\sigma + \chi B_y^2}. 
\eeq
This means that the angular dependence of the negative magnetoresistance signal is in fact 
stronger than implied by Eq.~\eqref{eq:57}. 
Instead of the $\cos^2 \theta$ dependence, we have a Lorentzian at small angles, with the angular width 
\beq
\label{eq:61}
\Delta \theta \sim \sqrt{\frac{\sigma}{\chi B^2}}.
\eeq
This angular narrowing of the negative magnetoresistance is in fact observed experimentally~\cite{Ong_anomaly,Li_anomaly}. 

Another important consequence of the anomalous density response, described by Eq.~\eqref{eq:40}, is the existence of 
nonlocal transport phenomena~\cite{Parameswaran14}, somewhat similar to the ones observed in graphene~\cite{Abanin11}. 
This also follows from the fact that the chiral charge diffusion length $L_c$ is a macroscopic length scale in Weyl metals. 
When $L_c$ is large, nonequilibrium chiral charge, created by current in the presence of an applied 
magnetic field in one part of the sample, will lead to voltage in response to an applied magnetic field 
far away from the current path. 
This is similar to the generation of voltage, transverse to the direction of the current, described by Eq.~\eqref{eq:59}, 
but in addition involves diffusion of the nonequilibrium chiral charge $n_c \sim \chi (\bE \cdot \bB)$ over macroscopic 
distances $\sim L_c$. 

\section{CONCLUSIONS AND OUTLOOK}
\label{sec:4}
We have reviewed the current understanding of bulk topological transport phenomena in Weyl metals, focusing 
on general principles rather than specifics of particular materials. 
Some important topics have been left out of this review, in particular the interplay of bulk and surface state (Fermi arc) transport~\cite{Moll16,Parameswaran15,Beenakker16}, transport in inhomogeneous Weyl metals, in which an extra contribution 
to the anomalous transport, related to the ``chiral magnetic field" (curl of the chiral gauge field) is predicted to exist~\cite{Niu13,Vozmediano15,Franz16,Grushin_PRX}, and type-II Weyl semimetals~\cite{Soluyanov15}. 

The significance of the transport phenomena in Weyl metals, related to the chiral anomaly, is that these represent new
observable macroscopic quantum phenomena.  
This is particularly clear from the highly nontrivial non-Ohmic scaling behavior of the conductance, described in Section~\ref{sec:3}. 
One may compare this to the nontrivial conductance scaling that arises due to quantum interference phenomena, which lead to 
Anderson localization~\cite{Lee85}, also a macroscopic quantum effect. 
An important difference however, is that, while localization phenomena can only be observed in small samples at very low 
temperatures, the anomalous transport phenomena in Weyl metals survive up to high temperatures, 
on the order of 100~K~\cite{Ong_anomaly,Li_anomaly}, and exist in macroscopic samples. 

An obvious question for the future work is whether there are other manifestations of the chiral anomaly, which go beyond 
transport phenomena, described in this article. 
A promising candidate is the magnetic response. This is likely to be nontrivial in Weyl semimetals due to both the 
Landau level structure, which will give rise to anomalies in orbital response and the spin-momentum locking, which is certain to affect the Pauli paramagnetic response. 
Interesting anomalies in magnetic response have in fact already been 
observed in one Weyl semimetal representative, NbAs~\cite{Analytis16}. 

Another area of interest is the interplay of electron-electron interactions and the nontrivial topology of the Weyl nodes. 
Some progress in this area has already been made. 
An incomplete list of topics that have been addressed are: the interplay of the chiral anomaly and spin and charge collective modes in Weyl metals~\cite{Liu13,Panfilov14,Xiao15,Nomura16}; Luttinger liquid physics within the lowest chiral Landau level~\cite{Nagaosa16}; potential strong correlation effects in Weyl semimetals~\cite{Nagaosa_WMI,Grushin16}; and hydrodynamics in ultra-clean Weyl metals~\cite{Sachdev16}. 
The interplay of superconductivity and topology of Weyl fermions is also of significant interest~\cite{Meng12,Moore12,Aji14,Tanaka15,Bednik15,YiLi15,Bednik16} due to potential for 
topological superconductivity with Majorana edge states. 
In particular, superconductivity in magnetic Weyl semimetals is necessarily topologically 
nontrivial~\cite{Meng12,Bednik15,YiLi15}, since the Berry curvature flux through the Fermi surface, felt by the 
Cooper pairs, forces point nodes to appear in the superconducting gap function~\cite{Bednik15,YiLi15}. 
The node projections on the surface BZ are then connected by Majorana arc edge states, in close analogy to how the Fermi arcs connect projections of the Weyl nodes. 

On the experimental front, there is a clear need for ``better" Weyl semimetal materials, which means smaller number of 
Weyl node pairs, Fermi energy closer to the nodes and no other states near the Fermi energy. 
As described above, universality of the chiral anomaly related response in Weyl and Dirac metals is an emergent property, which
becomes more precise as the energy is reduced towards the band-touching nodes. 
Currently, only Na$_3$Bi and ZrTe$_5$ come sufficiently close to the ideal limit of a topological semimetal to exhibit 
clear unequivocal transport signatures of the chiral anomaly, but this is certain to change in the near future as more 
topological semimetal materials are discovered. 

\section*{DISCLOSURE STATEMENT}
The authors are not aware of any affiliations, memberships, funding, or financial holdings that
might be perceived as affecting the objectivity of this review. 

\section*{ACKNOWLEDGMENTS}
We gratefully acknowledge Leon Balents, Grigory Bednik, Yige Chen, Yong Baek Kim, Ivan Panfilov, Dmytro Pesin and Alexander Zyuzin for collaboration on topics, covered in this review. We also acknowledge Claudia Felser, M. Zahid Hasan and Nai Phuan Ong
for numerous useful discussions. Financial support was provided by the Natural Sciences and Engineering Research Council (NSERC) of Canada. 
\bibliographystyle{ar-style4}
\bibliography{references}

\end{document}